\documentclass[a4paper,twocolumn,english,prl, showpacs]{revtex4}
\usepackage[T1]{fontenc}
\usepackage[latin2]{inputenc}
\usepackage{amsmath}
\usepackage{graphicx}
\usepackage{amssymb}

\makeatletter
\usepackage{graphicx}
\usepackage{graphicx}

\makeatother

\usepackage{babel}

\begin{document}

\title{Carbon nanotube quantum pumps}

\author{L. Oroszl\'any$^{1
}$, V. Z\'olyomi$^{1,2}$, and C. J. Lambert$^{1}$}

\affiliation{$^{1
}$Physics Department, Lancaster University, LA1 4YB, Lancaster,
United Kingdom}

\affiliation{$^{2
}$Research Institute for Solid State Physics and Optics of the
Hungarian Academy of Sciences, P. O. B. 49, H-1525, Budapest, Hungary}


\begin{abstract}
Recently  nanomechanical devices composed of a long stationary
inner carbon nanotube and a shorter, slowly-rotating outer tube
have been fabricated. In this Letter, we study the possibility of
using such devices as adiabatic quantum pumps. Using the Brouwer
formula, we employ a Green's function technique to determine the
pumped charge from one end of the inner tube to the other, driven
by the rotation of a chiral outer nanotube. We show that there is
virtually no pumping if the chiral angle of the two nanotubes is
the same, but for optimal chiralities the pumped charge can be a
significant fraction of a theoretical upper bound.

\end{abstract}
\maketitle \label{Introduction}Quantum pumps are time-dependent
electron scatterers, which are able to transport electrons between
two external reservoirs. They are adiabatic if the frequency of
the pump cycle is smaller than the inverse of the characteristic
timescale of the scatterer, namely the Wigner delay time
\citep{WignerDelay_1955}. Recent experimental
\citep{Pothier1992-SEQP,SwitkesM_1999_1} and theoretical
\citep{ZhouF_1999_1,AvronJE_2000_1,AvronJE_2001_1,wei-wang-cnt-pump,
MoskaletsM_2002_1,governale-interacting-pump,TorresLEFF_2009_1}
studies of adiabatic quantum pumps have examined the conditions
for optimal pumping and the effects of noise and dissipation. All
of these devices are based on electrical pumping. In this work, we
propose and analyze a novel realization of a mechanically-driven
quantum pump.

The significance of mechanically-driven quantum pumps lies in
their ability to convert mechanical energy to electrical energy,
which could be used for energy scavenging, via the conversion of
ambient vibrational energy to electrical energy (see for example
Ref. \citep{RoundyS_2003_1}). The pumped current could be used to
power or control nanoscale electronic devices, making it a useful
component in NEMS devices. As it will be shown below, the proposed
nanomechanical pump can operate at 30-40\% of the theoretical
upper limit, which makes it highly attractive as an energy
scavenger.

Our analysis was stimulated by recent experiments
\citep{FennimoreAM_2003_1,BarreiroA_2008_1}, which demonstrate
that it is possible to engineer a double-walled carbon nanotube in
such a way that the inner tube is fixed, and the outer is caused
to rotate around it by an external force. In this paper we
demonstrate that such a device can also be used as a quantum pump.
The basic idea is that if the two nanotubes have different
chirality, the rotation of one of the tubes will produce a
time-dependent potential that induces electron flow in the other.
Such flow is clearly allowed by symmetry, but the question of
whether or not the pumped charge is significant must be answered
by a quantitative calculations based on a realistic Hamiltonian.
In what follows the results of such a calculation are presented.

\begin{figure}
\includegraphics[width=0.7\columnwidth]{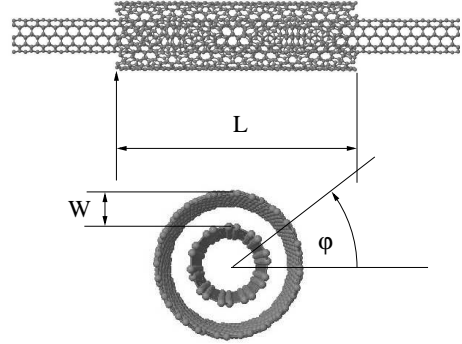}

\caption{\label{fig:shuttle} The shuttle geometry used throughout
our calculations. An outer nanotube of length L concentrically
surrounds an inner tube of length 2L, with an inter-layer spacing
W corresponding to the van der Waals distance ($L\approx50$~\AA,
$W\approx3.4$~\AA). The inner wall remains fixed, while the outer
tube is rotated about the tube axis.}

\end{figure}

We calculate the adiabatically-pumped charge in the double-walled,
carbon-nanotube, shuttle geometry shown in Figure
\ref{fig:shuttle}, which mimicks the experimental setup of Ref.
\citep{FennimoreAM_2003_1}. The inner tube is fixed, while the
shorter outer tube slowly rotates. The adiabatic charge pumped by
a time-varying scatterer connected to external reservoirs by
scattering channels (labelled $j$) is given by the Brouwer formula
\citep{ButtikerM_1994_1,BrouwerPW_1998_1}, which states that the
pumped charge $Q_{j}$ in the $j$th channel is given by
$\dot{Q_{j}}\approx\left(e/h\right)E_{jj}$, where $E_{jj}$ is the
energy shift matrix as defined by
$E\left(t,\mu\right)=i\hbar\partial_{t}S\left(t,\mu\right)S^{\dagger}\left(t,\mu\right)$,
$S$ is the scattering matrix, and $\mu$ is the Fermi energy. In
what follows, the Hamiltonian used to build the $S$ matrix is
constructed from the inter-molecular H\"uckel model (IMH), which
is a tight binding model with inter-molecular interactions
determined by the geometrical arrangements of atoms within a
device \citep{StafstromIJQC,LaplacePRA,ZACHARY_2006_1}. For a
given pair of inner and outer carbon nanotubes, the Green's
function and scattering matrix are determined from the IMH
Hamiltonian via Dyson's equation
\citep{rocha:085414,DattaMesobook_1995_1}. Brouwer's formula is
evaluated from appropriate derivatives of scattering matrix
elements. To reveal the rich behavior of this family of quantum
pumps, results are obtained for different choices of the Fermi
energy (measured in units of the $\gamma$ nearest neighbor
intramolecular hopping matrix element).

\begin{figure}
\includegraphics[width=0.7\columnwidth]{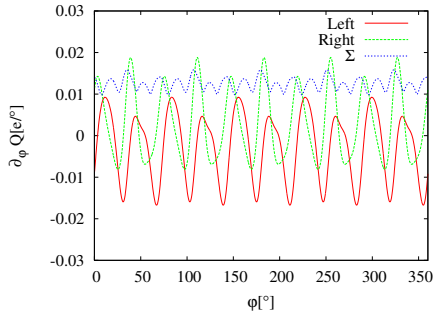}

\caption{\label{fig:paremplot} (color online) The calculated
parametric emissivity in a (5,5)@(15,5) shuttle pump at a Fermi
energy of $0.0081\gamma$ as a function of the rotational angle
$\varphi$ (solid: charge pumped left; dashed: charge pumped
right). $\Sigma$ marks the average of the magnitude of the
off-diagonal elements of the $E_{jj}$ matrix, which should be zero
for an optimal quantum pump \citep{AvronJE_2001_1}. }

\end{figure}

We focus on the (5,5) and (9,0) inner nanotubes with several
different outer tubes which were chosen such that the inter-layer
distance would roughly correspond to the van der Waals distance.
Figure \ref{fig:paremplot} shows the parametric emissivity (ie.
the trace of the energy shift matrix), as a function of the
rotational angle $\varphi$ for the (5,5)@(15,5) shuttle pump.
Depending on the particular angle, charge may be pumped either
from left to right or vice versa. The integral of this parametric
emissivity within a full cycle is the number of pumped electrons
per cycle. The length of the cycle is determined by the rotational
symmetry of the inner nanotube; in the case of the (5,5), there is
a C$_{5}$ rotational symmetry, hence the cycle is from 0$^{\circ}$
to 72$^{\circ}$. This can be clearly seen in Figure
\ref{fig:paremplot}, as the plot is periodic with a period length
of 72$^{\circ}$. According to Ref. \citep{AvronJE_2001_1}, a
quantum pump is optimal if the off-diagonal elements of the energy
shift matrix are zero. We may define $\Sigma$ as the average of
the absolute values of the off-diagonal elements, which can be
interpreted as a measure of the deviation from optimal behavior.
Results for $\Sigma$ are shown in Figure \ref{fig:paremplot},
which demonstrates that the pumping is not optimal. However, as we
will see below, at low Fermi energies and near Fabry-Perot
resonances the pumping can be quite high and approaches a
significant fraction of the theoretical limit.

\begin{figure}
\includegraphics[width=0.7\columnwidth]{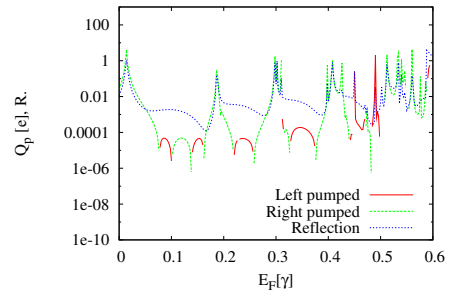}

\caption{\label{fig:14-6plot} (color online) The calculated pumped
charge per 360$^{\circ}$ rotation in a (5,5)@(14,6) shuttle pump
on a logarithmic scale as a function of the shift of the Fermi
energy (green dashed: charge pumped left; red solid: charge pumped
right). At certain energies, the pumped charge is very high. These
peaks correlate with the Fabry-Perot resonances in the reflection
coefficient (blue dotted).}

\end{figure}

In Figure \ref{fig:14-6plot} we show the pumped charge in a (5,5)
carbon nanotube with a (14,6) outer nanotube slowly rotating
around it. The charge pumped per 360$^{\circ}$ rotation is
obtained by calculating the parametric emissivity from the left to
the right lead and vice versa at different angles, and then
integrating the result from 0$^{\circ}$ to 360$^{\circ}$.
Continuity is satisfied, because the charge pumped into the right
lead equals the charge taken from the left lead, with high
numerical accuracy. When the time derivative of the S matrix is
small, this requires a fine integration mesh.

The average pumped charge clearly drops by several orders of
magnitude as the Fermi energy is increased, thus for the most
efficient pumping the Fermi energy should be close to the Dirac
point. Note however, that the pumped charge could again increase
if the Fermi level is large enough to open another channel. Beyond
this average behavior, it is also important to note the presence
of numerous sharp peaks in the pumped charge. The location of
these peaks correlates with the Fabry-Perot resonances in the
reflection coefficient. In other words, when the transmission is
high, pumping is low, and vice versa. Indeed, when the coupling
between the shells is strong, the associated increase in
scattering leads to decreased transmission and increased
reflection, while at the same time the pumped charge is increased.
This suggests that the largest pumping occurs at Fabry-Perot
resonances. However, the location of these peaks in energy is very
sensitive to the geometry and strongly depends on the length of
the outer tube and the structure of the edge of the tube.
Therefore in an actual experiment, these features will likely be
averaged out. For this reason, the most efficient pumping will be
at low energies where the average pumped charge is highest (or
near an energy where the next channel opens).

A further noteworthy feature of the quantum pumps studied here is
that the direction of the pumped charge changes sign at certain energies.
While this could in principle be used to change the direction of the
current by shifting the Fermi energy while maintaining the rotation
of the outer shell in the same direction, experimentally it is difficult
to achieve this. Furthermore in the region where the sign change
takes place, the charge pumping is already at least an order of magnitude
smaller than at low doping levels, which would further hinder such
applications.

\begin{figure}
\includegraphics[width=0.8\columnwidth]{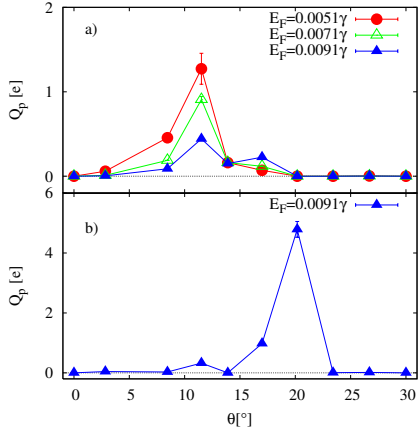}

\caption{\label{fig:efficiency} (color online) The average pumped charge in
a (5,5) ($a$) and a (9,0) ($b$) carbon nanotube as a function of
the chiral angle of the outer tube. The averaging is performed over
different relative positions along the tube axis. The most efficient
chiralities are the ones around the chiral angle of $\approx10^{\circ}$
(where $0^{\circ}$ corresponds to the zigzag tubes) for the (5,5)
and near $\approx20^{\circ}$ for the (9,0). There is practically
no pumping if the outer tube is achiral, as expected.}

\end{figure}

To identify which chirality has the highest efficiency, Figure
\ref{fig:efficiency} shows the pumped charge per 360$^{\circ}$
rotation in the (5,5) and the (9,0) inner tubes with different
outer nanotubes, for a number of different energies and chiral
angles of the outer tube (averaging is performed over relative
positions of the two shells along the tube axis, see below). In
the case of the (5,5) there are a few chiralities near the angle
$\approx10^{\circ}$ (where $0^{\circ}$ corresponds to the zigzag
tubes) where the pumping is very high, although it doesn't reach
the theoretical limit (see discussion below). Tubes of such
chirality are efficiently pumping electrons through a (5,5) inner
tube. On the other hand, neither the (10,10), nor the (18,0)
produces any significant charge pumping. This is expected, since
the pumping is unlikely to occur if the chiral angles are the same
or if both the inner and outer tubes are achiral.

An important question is whether or not the position
of the outer shell along the tube axis with respect to the inner tube
has significant effect on the pumped charge. We performed calculations
to check the magnitude of this effect, calculating the pumped charge
at 10 different inequivalent positions. The pumped charges in Figure
\ref{fig:efficiency} are obtained from the average of these calculations,
and the plotted errorbar shows how much these values vary. This demonstrates
that translating the outer tube relative to the inner tube produces only
a small change in the pumped charge. (Note however, that the locations
of the aforementioned sharp peaks associated with the Fabry-Perot
resonances in the reflection coefficient are sensitive to such effects.)

We have also calculated the pumped charge in a (9,0) inner tube
using the same outer tubes (see bottom half of Figure
\ref{fig:efficiency}). These results are similar to those of the
(5,5) except that the plot has a peak at around
$\approx20^{\circ}$. This result suggests that the optimal chiral
angles are such that the difference of the chiral angle of the
inner and outer shell is $\approx20^{\circ}$. (A similar result
was found for the optimal momentum transfer between two nanotubes
in the so-called carbon nanotube windmill, which is essentially
the inverse of the effect studied here. \citep{BaileySWD_2008_1})
A further difference between the results on (5,5) and (9,0) is the
vertical scale: the maximum pumped charge is larger in the latter.
This is because the integration is performed on a 360$^{\circ}$
interval, which contains 5 parametric cycles in the case of (5,5)
and 9 cycles in the case of (9,0). This suggests that from a
practical point of view, the best inner shells for use in a carbon
nanotube quantum pump are the ones with high rotational symmetry.
According to Ref. \citep{AvronJE_2001_1}, the maximum pumped
charge per parametric cycle is one per channel, so the theoretical maximum
for a 360$^{\circ}$ rotation in the (9,0) at low energies (where
there are 2 open channels) is 18. The highest pumping found in our
calculations is approximately one third of this.
\begin{figure}
\includegraphics[angle=270,width=0.9\columnwidth]{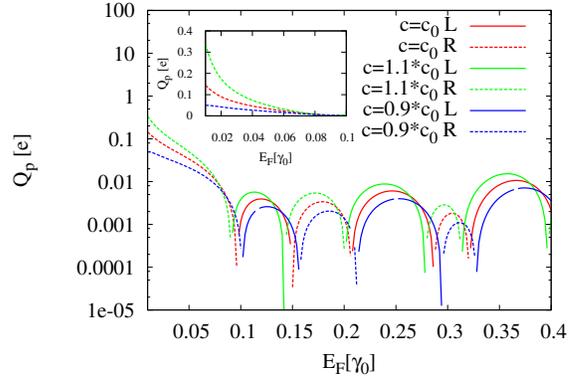}

\caption{\label{fig:coupling} (color online) The charge pumping in
the (5,5)@(15,5) shuttle pump for different values of the strength
of the inter-layer coupling (solid: charge pumped left; dashed:
charge pumped right). A $\pm10\%$ change in the coupling alters
the pumped charge roughly by a factor 2-3, resulting in nearly an
order of magnitude difference when comparing the cases of the
$10\%$ weakened and $10\%$ strengthened coupling.}

\end{figure}
The IMH tight-binding Hamiltonian used in the calculations,
utilizes inter-layer interactions which were fitted to the Davydov
splitting of ethylene \citep{StafstromIJQC}. This model was
recently demonstrated to predict charge transfer in double-walled
carbon nanotubes \citep{ZACHARY_2006_1} which agrees well with
experiments \citep{Hide_2006_1,Hide_2006_2}. Nevertheless it may
be possible that a slightly different inter-layer coupling can
provide more accurate results. For this reason we have examined
the effect of slightly altering the inter-layer coupling strength.
Figure \ref{fig:coupling} shows that the strength of this coupling
significantly influences the pumped charge. Changing the magnitude
of the coupling by around $10\%$ yields nearly a factor 2-3 change
in the pumped charge at low energies. This effect could therefore
be exploited to probe the strength of the inter-layer coupling.
Indeed by measuring the pumped charge in the proposed geometry, it
may be possible to fit the IMH parameters directly to
measurements.

\label{Conclusion}In conclusion, we have studied the pumped charge
in a double-walled carbon-nanotube shuttle geometry, consisting of
a long (5,5) or (9,0) inner tube and a short outer shell of
varying chirality. We have demonstrated that charge pumping can be
a significant fraction of a theoretical upper bound and that the
most efficient pumping occurs when the inner tube has a high
rotational symmetry around the tube axis and the difference in the
chiral angle of the two shells is $\approx20^{\circ}$. We have
also found that the pumped charge is sensitive to the inter-shell
coupling in the system. Our aim has been to provide a first
demonstration of significant pumping in such devices and therefore
we have focused on clean nanotubes in the adiabatic limit.  For
the future it will be  of interest to consider the effects of
disorder and non-adiabaticity. Regarding the former, one notes
that at least in one dimension, disorder which preserves the
spatial symmetry of a lattice does not completely randomize the
phase of scattering states \cite{lamb1,lamb2} and therefore phase
derivatives, which are at the heart of the Brouwer formula can be
expected to retain a memory of the underlying chirality.
Furthermore in the absence of commensurability, translating the
outer tube relative to the inner tube induces a range of different
incommensurate scattering potentials and as shown in Figure \ref{fig:efficiency},
this does not destroy charge pumping. Regarding the question of
non-adiabaticity, our results are valid as long as the frequency
of rotation is smaller than the inverse of the Wigner delay
time \citep{WignerDelay_1955}, which for the nanotubes we have
studied is on the order of $10^{-11}$ seconds near the resonances
and even smaller, $10^{-14}$ seconds far from resonances. This
suggests that the Wigner delay does not raise any technical
barriers before the realization of adiabatic pumping and more
likely electron-phonon coupling will set an upper bound to the
operating frequency \citep{ServantieJ_2006_1}.

\begin{acknowledgments}
\label{ack:Support-from-OTKA}Support from OTKA in Hungary (Grants
No. F68852, and K60576) and the EPSRC in the UK is gratefully
acknowledged. V. Z. also acknowledges the J\'anos Bolyai Research
Foundation of the Hungarian Academy of Sciences and the Marie
Curie IEF projects FUNMOLS, NANOCTM and NANOTRAN
(PIEF-GA-2008-220094). We thank Dr. P. R. Surj\'an for valuable
discussions.
\end{acknowledgments}

\end{document}